%% file: paper.tex
\documentclass[sigconf]{acmart}

\usepackage{booktabs} 
\usepackage{subcaption}

\setcopyright{rightsretained}
\acmConference[SIGIR 2018 eCom]{ACM SIGIR Workshop on eCommerce}{July 2018}{Ann Arbor, Michigan, USA} 
\acmYear{2018}
\copyrightyear{2018}

\begin{document}
\title{Did We Get It Right? Predicting Query Performance in E-commerce Search}

\author{Rohan Kumar}
\orcid{1234-5678-9012}
\affiliation{%
  \institution{Flipkart}
}
\email{rohankumar@flipkart.com}

\author{Mohit Kumar}
\affiliation{%
  \institution{Flipkart}
}
\email{k.mohit@flipkart.com}

\author{Neil Shah\footnote{Dr. Shah is now at Snap Inc.}}
\affiliation{%
  \institution{Carnegie Mellon University}
}
\email{neilshah@cs.cmu.edu}

\author{Christos Faloutsos}
\affiliation{%
  \institution{Carnegie Mellon University}
}
\email{christos@cs.cmu.edu}






\begin{abstract}

In this paper, we address the problem of evaluating whether results served by an e-commerce search engine for a query are good or not. This is a critical question in evaluating any e-commerce search engine. While this question is traditionally answered using simple metrics like query click-through rate (CTR), we observe that in e-commerce search, such metrics can be misleading. Upon inspection, we find cases where CTR is high but the results are poor and vice versa. Similar cases exist for other metrics like time to click which are often also used for evaluating search engines.

We aim to learn the quality of the results served by the search engine based on \emph{users' interactions with the results}. Although this problem has been studied in the web search context, this is the first study for e-commerce search, to the best of our knowledge. Despite certain commonalities with evaluating web search engines, there are several major differences such as underlying reasons for search failure, and availability of rich user interaction data with products (e.g. adding a product to the cart). We study \emph{large-scale} user interaction logs from Flipkart's\footnote{Flipkart is the largest e-commerce platform in India.} search engine, analyze behavioral patterns and build models to classify queries based on user behavior signals. We demonstrate the feasibility and efficacy of such models in accurately predicting query performance. Our classifier is able to achieve an average AUC of 0.75 on a held-out test set.

\end{abstract}

%
%



\maketitle

\section{Introduction}

\input{1-introduction}

\section{Related Work}
\input{2-related-work}

\section{Query performance judgements}
\input{3-dataset-description}


\section{Experiments}
\input{5-experiments}

\section{Conclusion And Future Work}
\input{6-conclusion}

\section{Acknowledgements}
\input{7-acknowledgements}

\bibliographystyle{ACM-Reference-Format}
\bibliography{sample-bibliography} 

\end{document}

%% file: 1-introduction.tex
\begin{figure}
\centering
\includegraphics[width=.5\columnwidth]{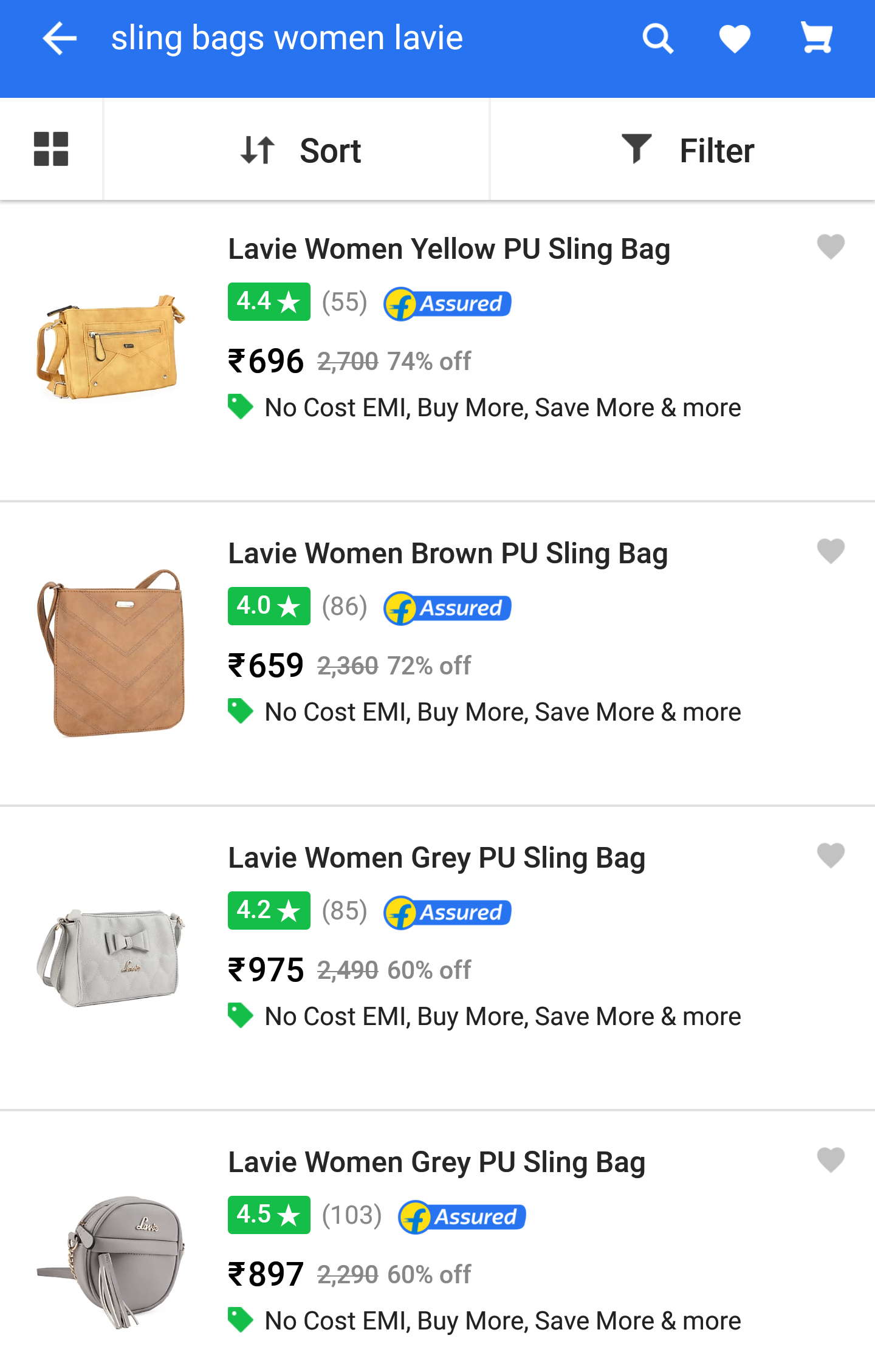}
\vspace{-0.5\baselineskip}
\caption{Mobile app e-commerce results page for the query ``sling bags women lavie'', showing relevant products.}
\label{fig:query_results}
\end{figure}

Search engines are a fundamental component of most modern Internet applications, and evaluating their performance on a query is not only needed for evaluating their overall performance, but is also critical in the iterative process of improving the algorithms that power them. This is important since bad performance of a search engine leads to customer attrition as described in \citet{white2009characterizing}. Traditionally, the performance of a search engine on a query is measured using metrics derived from ordinal ratings of the search results given by human experts \cite{chapelle2009expected, jarvelin2002cumulated, zhou2007query}. However, obtaining such manual judgments is prohibitive for the large document collections and high number of unique queries commonly encountered in most modern Internet applications. While one could solicit explicit feedback on the quality of search results from the users of a search engine, this may be detrimental to their experience of the application.

More recent work \cite{guo2010predicting} has focused on automating the evaluation of search engine performance by using implicit feedback on the quality of search results derived from various user activity signals generated by the interactions between users and the results presented to them. Most of this work has been done for Internet search engines while in this paper, we focus on e-commerce search engines. The users of e-commerce applications tend to look for products and services, and thus the queries typically encountered by e-commerce search engines are fundamentally different from the informational and navigational queries typically encountered by Internet search engines.

The most popular user activity signal in the aforementioned work is clicks and it is used to define the Click-Through Rate (CTR) metric. The CTR of a query is often used as a proxy for the performance of the search engine on that query, and this approximation is based on the assumption that clicks on search results are a reliable indicator of performance. However, \citet{beyondclicksCIKM13} points out that while clicks are a useful indicator of performance, they can nevertheless be quite noisy.

\begin{figure}
\centering
\includegraphics[width=\columnwidth]{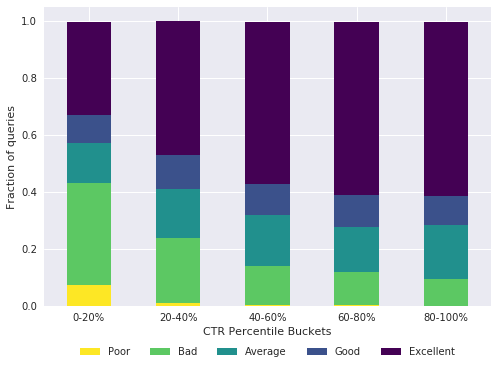}
\caption{Distributions of search result ratings across percentile based CTR buckets.}
\label{fig:ctr_buckets}
\end{figure}

We validate this observation for e-commerce search by studying the distributions of the ordinal ratings of search results given by human experts to queries having a wide range of CTR values randomly sampled from Flipkart search query-logs. We discretized the CTR values into $5$ buckets with the bucket boundaries at the $20$th, $40$th, $60$th, and $80$th percentiles of the CTR values of our sampled queries. The distributions of search result ratings across these percentile-based CTR buckets is shown in Figure~\ref{fig:ctr_buckets}. The details of how queries are sampled from our query-logs and how the associated search results are rated by human experts are given in Section~\ref{section:dataset}.

From Figure~\ref{fig:ctr_buckets} it is evident that while the fraction of queries whose results are rated as poor decreases as we go from the lowest CTR bucket to the highest CTR bucket, a significant fraction of queries whose results are rated as bad still exists even in the highest CTR bucket. Figure \ref{fig:query_results} shows an example of a search engine results page (SERP) that appears in Flipkart's mobile app for the query ``sling bags women lavie''. The query has good results even though it belongs to the 0-20\% CTR bucket from Figure \ref{fig:ctr_buckets}. This highlights the need for a richer set of user activity signals beyond click behavior. \citet{guo2010predicting} made use of such signals, but their focus was on Internet search where the set of user activity signals available is limited in comparison to e-commerce search, where we have additional signals available such as the time taken to click an add-to-cart or buy-now button. Using a richer set of such user activity signals, we build a classification model to predict whether the results for any query from our query-logs would be rated as bad or good by human experts and thus automate the evaluation of our search engine performance. Such a system also serves as a first step towards building a system to predict user satisfaction at the level of individual user activity sessions as studied in \cite{fox2005evaluating, beyonddcgWSDM10, strugglingSusanCIKM15}.

Our classifier is able to achieve an average AUC of $0.75$ on a held-out test set. On certain product categories like Mobile Phones, we achieve an average AUC of $0.88$ on the held-out test set.

Summarily, the primary contributions of our work are:
\begin{enumerate}
\item We identify a rich set of user activity signals that help predict whether the results for any search query would be rated as bad or good by human experts.
\item We demonstrate that it is possible to use user activity signals to automate the evaluation of search engine performance for e-commerce applications.
\item We analyze the performance of our classifier and derive insights into the effectiveness of automated systems for evaluating search engine performance that are of particular interest to e-commerce applications.
\end{enumerate}

%% file: 2-related-work.tex


\subsection{Query Performance}
Evaluating search engine performance has been well-studied in the domain of web search. Topical relevance based metrics like nDCG \cite{jarvelin2002cumulated}, expected reciprocal rank \cite{chapelle2009expected} and weighted information gain \cite{zhou2007query} require explicit human labeled relevance judgments for query-document pairs which are prohibitively costly to calculate at scale for real-world web scale evaluation.

Several methods were proposed to automatically measure various characteristics of the documents retrieved for a query, which can then be used for measuring overall system performance. Clarity score \cite{cronen2002predicting} evaluates query performance by measuring the relative entropy between query language model and corresponding collection language model. The Robustness score \cite{zhou2006ranking} exploits the fact that query-level ranking robustness is correlated with retrieval performance. It is measured as the expected value of Spearman\textquotesingle s rho between ranked lists from original collection and a corrupted collection. \citet{carmel2006makes} find Jensen-Shannon divergence between queries, relevant documents and the entire collection to be an indicator of query performance. However, \cite{zhou2007query} experimentally show the ineffectiveness of these metrics in measuring search performance on web-scale engines.

User click behavior has been used as an alternative to expert judgments for automatically tuning retrieval algorithms (predicting document relevance) as well as estimating IR evaluation metrics \cite{joachims2005accurately, carterette2008evaluating, cikm2009tut, guo2010predicting}. \citet{dwellWSDM14} show that only analysing user clicks naively may not indicate satisfaction, but rather using dwell time per click appropriately indicates query level satisfaction in a better manner. \citet{guo2010predicting} also make use of interaction features and engine switches as signals to predict DCG@3.


\subsection {Search Session Performance}
There has been considerable work in the area of analyzing user satisfaction at a session level rather than at an individual query level. \citet{fox2005evaluating} conducted one of the first studies that found association between explicit ratings and implicit measures of user interest, concluding that user satisfaction can be predicted using such implicit signals. \citet{beyonddcgWSDM10} show empirically that user behavior alone can give an accurate picture of the success of the user\textquotesingle s web search goals, without considering the relevance of the documents displayed. There have been studies focusing on graded satisfaction \cite{gradedWSDM15} as well as specific user behaviors like query reformulation \cite{beyondclicksCIKM13,strugglingSusanCIKM15} and interaction sequences \cite{interactionSIGI17} for understanding satisfaction.

\subsection {E-commerce Search Performance}
Most studies have been geared towards web search where user search goals are different from those in product/e-commerce search. However, there has been some work recently in the context of product search. \citet{ebaySIGIR11} study the user behavior in the e-commerce search context in a specific scenario when the search engine doesn't retrieve any results. \cite{productWSDM18} is a recent study that addresses the user's session satisfaction in product search. They approach the problem by firstly identifying a taxonomy of user intents while interacting with product search, and then analyze the user's behavior in the context of the defined taxonomy. They predict user session satisfaction by utilizing the interaction behavior, where they build separate models for different intents with the demonstration that user behavior is different under different intents. 
Our work, while building upon the learnings from these studies, differs in that we are interested in measuring only the aggregate query performance instead of more user-centric task of session satisfaction. The example mentioned by \citet{productWSDM18} where the results expected by two different users for the same query \emph{iphone} may be different and thus they may be individually dissatisfied even though the results shown are ``relevant.'' We aim to address the simpler, albeit more business-critical problem of understanding a query's result relevance in a user-agnostic fashion. The underlying reason(s) for a search engine's poor query performance is due to factors like incorrect spell error handling, vocabulary gap \cite{carpineto2012survey}, selection gap (when the e-commerce platform does not sell a particular item -- e.g. \emph{chocolate} when packaged food items are not sold), and more. Thus understanding and measuring the user-agnostic query performance can help improve the core relevance algorithm of the search engine. 



%% file: 3-dataset-description.tex
\label{section:dataset}



At Flipkart, regular search quality analysis is done for a random sample of queries (stratified on query volume segment) from search logs by a team of quality experts. They are requested to rate queries on a five-point scale (PBAGE: Poor-1, Bad-2, Average-3, Good-4, Excellent-5) based on result relevance. To ensure the consistency of labeling across experts, inter-rater agreement is continuously monitored. In this work, we make use of the expert editorial judgments for the month of January 2018.


We selected 18,613 queries from this randomized set of expert-labeled queries which occurred more than 100 times in a week in order to ensure reasonable user activity data. This set of queries corresponded to 127M query impressions, 149M clicks and 14M other interactions (e.g. filters application, sort application) from activity by 21M users collectively spending almost 4M hours on the platform. The data is collected from Flipkart's mobile app, significantly reducing the chances of bot traffic. All user behavior data is captured for the same week in which the query was labeled by an expert. We assume the search system and hence user activity remain constant throughout the week as there are no manual or algorithmic fixes applied during the week.

\begin{figure*} [t]
  \centering
  \begin{subfigure} {.24\textwidth}
  \includegraphics[width=\linewidth]{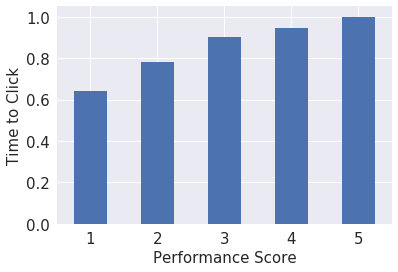}
  \caption{Time to first click}
  \label{timeToFirstClick_mean}
  \end{subfigure} \hfill
  \begin{subfigure} {.24\textwidth}
  \includegraphics[width=\linewidth]{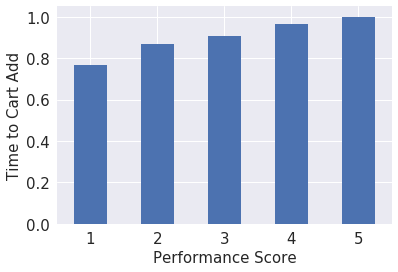}
  \caption{Time to first cart}
  \label{timeToFirstCart_mean}
  \end{subfigure} \hfill
  \begin{subfigure} {.24\textwidth}
  \includegraphics[width=\linewidth]{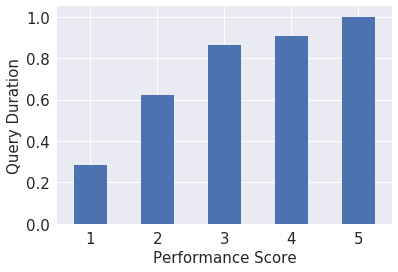}
  \caption{Query duration}
  \label{queryDuration_mean}
  \end{subfigure} \hfill
  \begin{subfigure} {.24\textwidth}
  \includegraphics[width=\linewidth]{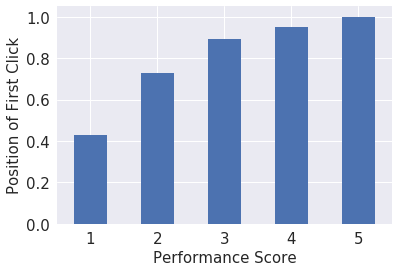}
  \caption{First Click Position}
  \label{firstClickPos_mean}
  \end{subfigure}
  \vspace{-0.5\baselineskip}
  \caption{Normalized Distribution of Activity Time and Positional feature values with respect to Performance Score}
  \label{fig:activitytime_positional}
\end{figure*}

\section{Signals of user behavior}
\label{section:signals}

\begin{table}
\centering
\caption{Features used to distinguish between good and poorly performing search queries}
\label{features}
\begin{tabular}{|l|l|}
\hline
\multicolumn{2}{|c|}{\textbf{Activity time}}                                                                                \\ \hline
timeToFirstClick & \begin{tabular}[c]{@{}l@{}}Time taken to click first product\end{tabular}           \\ \hline
timeToFirstCart  & \begin{tabular}[c]{@{}l@{}}Time taken to add a product to the cart\end{tabular} \\ \hline
queryDuration    & \begin{tabular}[c]{@{}l@{}}Total dwell time of the query\end{tabular}                            \\ \hline
\multicolumn{2}{|c|}{\textbf{Positional}}                                                                                \\ \hline
posFirstClick    & Position of first product clicked                                                                              \\ \hline
\multicolumn{2}{|c|}{\textbf{Activity aggregates}}                                                                               \\ \hline
numClicks        & Number of clicks                                                                                               \\ \hline
numSwipes        & Number of swipes                                                                             \\ 
\hline
numCarts   & Number of cart adds                               \\ \hline
numFilters       & Number of times a filter was applied                                                                           \\ \hline
numSorts         & Number of times user changed sorting                                                                           \\ \hline
numImpressions   & \begin{tabular}[c]{@{}l@{}}Number of product impressions \\ in the viewport\end{tabular}           \\ \hline
clickSuccess   & Any product clicked for query                                \\ \hline
cartSuccess   & Any product added to cart for query
\\ \hline
\multicolumn{2}{|c|}{\textbf{Query text characteristics}}                                                                               \\ \hline
charQueryLen       & Length of the query in characters                                                                           \\ \hline
wordQueryLen         & Length of the query in words                \\ \hline
                                                               
LMScore         & Query language model perplexity score

\\ \hline
querySim         & Similarity to next query
\\ \hline
containsSP         & Query contains specifiers
\\ \hline
containsMT         & Query contains modifiers
\\ \hline
containsRS         & Query contains range specifiers
\\ \hline
containsUnits         & Query contains units like liters
\\ \hline
\multicolumn{2}{|c|}{\textbf{Meta aspects}}                                                                               \\ \hline
queryCat       &  \begin{tabular}[c]{@{}l@{}}Category (mobile phones, books etc.) \\of the query based on taxonomy \end{tabular}         
\\ \hline
queryType       & \begin{tabular}[c]{@{}l@{}}Type of the query (specific product, \\broad category etc.) \end{tabular}
\\ \hline
queryCount   & Frequency of the query                       \\ \hline
isAutoSuggestUsed    &  Auto-completed query or not         \\ \hline
isGoodNetwork       & Network type is WiFi or 4G           \\ \hline
numProductsFound & Number of products matching the query   \\ \hline
\end{tabular}
\end{table}
Table \ref{features} lists the metrics along with their descriptions that we extracted for every query instance. We characterize the user behavior metrics as \textit{Activity time}, \textit{Positional} and \textit{Activity aggregates}. We characterize the non-user metrics as \textit{Query text characteristics} and \textit{Meta aspects}. 

\textit{Activity time} features capture the time taken by the user for various activities. timeToFirstClick is the time taken by the user to click a product after the results are displayes. timeToFirstCart is similar to timeToFirstClick except it captures time taken to add a product to the cart. queryDuration is the total time spent in interacting with the query results including all interactions with product pages, cart etc. Figures \ref{timeToFirstClick_mean}-\ref{queryDuration_mean} show the distribution of \emph{Activity time} features with respect to query performance. We observe interestingly that time taken for first click increases with the query performance. This is counter-intuitive in that when the query performance is good, it still takes users longer to click. This is however potentially explained with Figure \ref{numClicks_mean}, which shows the distribution of the number of clicks against query performance. We observe that when the query performance is low the total number of clicks is lower and it increases with query performance. Intuitively, the users usually don't click any products when the query performance is poor but when they click products for a poorly performing query they do it faster. Similar pattern is observed for the add-to-cart behavior, in Figures \ref{timeToFirstCart_mean} and \ref{numCarts_mean}.

\textit{Positional} features correspond to the position of result interaction. posFirstClick captures which position the user clicked first. A lower position value indicates that the results were shown near the top of the page. We observe that the average position of the first result click increases with improving query performance. This is correlated with the previous observation where time to first click of poorly performing queries is lower and correspondingly the user is clicking the results in lower positions (faster). The total number of clicks is low when query performance is low. Similar to Activity Time features, users usually don't click products when the query performance is poor but then they click products for a poorly performing query they do it at lower positions.

\textit{Activity aggregates} features capture the aggregated summary of user's actions for a query. We observe that all the activity aggregates are positively correlated with the query performance  -- i.e. increasing user activity indicates better query performance. Number of product clicks (numClicks: Figure \ref{numClicks_mean}), product swipes (numSwipes: Figure \ref{numSwipes_mean}), cart additions (numCarts: Figure \ref{numCarts_mean}), filters applied (numFilters: Figure \ref{numFilters_mean}), sort applied (numSorts: Figure \ref{numSorts_mean}), product impressions per query (numImpressions: Figure \ref{numImpressions_mean}), query successful click through rate (clickSuccess: Figure \ref{ctr_mean}), query successful cart conversion rate (cartSuccess: Figure \ref{cvr_mean}) are all positively correlated with query performance.

\textit{Query text characteristics} features capture the textual properties of the query. charQueryLen and wordQueryLen are length of query in characters and words respectively. LMScore is the perplexity score of the query based on a language model\cite{heafield2011kenlm} trained on the query logs. querySim is the text similarity between the current query and the following query defined by the measure described in \citet{hassan2014struggling}. We also make use of certain domain-dependent text features indicating if the query contains specifiers (e.g. ``greater than''), modifier phrases (e.g. ``least expensive''), range specifiers (e.g. ``between'') or units (e.g. ``liters'', ``gb''). The intuition here is that search engines may face difficulty in product retrieval when queries contain such phrases which require semantic understanding.

\textit{Meta aspect} features include additional information about the query. queryCat indicates the e-commerce product category. These are broad lines of business, namely Mobile Phones, Books, Electronics, Lifestyle, and Home and Furniture. Each query is assumed to belong to one of these categories. The intuition for using this feature is that the query performance and user behavior might be dependent on the specific categories. queryType indicates the type of query which is classified amongst three kinds, namely ``Product'', ``FacetCategory'
and ``Category.'' Queries in which the exact product that the user is looking for is mentioned are called ``Product'' queries (e.g. \emph{iPhone X}). Queries which refer to a broad group of products are called ``Category'' queries (e.g. \emph{shoes}). ``FacetCategory'' queries typically contain one or more attributes followed by a category (e.g. \emph{red Nike shoes}). For both queryCat and queryType, we make use of modules which are able to assign appropriate values for a given query (details of these modules is outside the scope of this paper).
queryCount is the total number of times the query was issued by users in the past week. isAutoSuggestUsed indicates whether the user selected the query from the suggested queries (auto-suggest). The intuition is that the queries suggested by the search engine typically perform better than query issued by user. isGoodNetwork indicates whether the user has a good Internet connection (defined as WiFi or LTE) while issuing the query. This is important, as the user experience and behavior might be altered if he/she doesn't have a good Internet connection leading to bad experience independent of the search engine's performance. numProductsFound indicates the total number of products found in the search index for the query. The intuition here is that the number of products found in conjunction with the type of the query may indicate if the search engine is not able to retrieve relevant results.

%% file: 5-experiments.tex
\begin{figure*} [t]
  \centering
  \begin{subfigure} {.24\textwidth}
  \includegraphics[width=\linewidth]{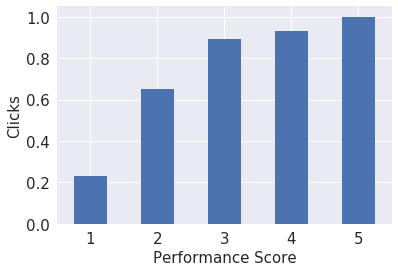}
  \caption{Number of clicks}
  \label{numClicks_mean}
  \end{subfigure} \hfill
  \begin{subfigure} {.24\textwidth}
  \includegraphics[width=\linewidth]{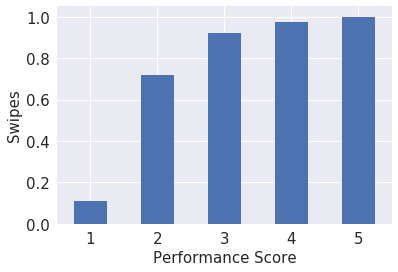}
  \caption{Number of swipes}
  \label{numSwipes_mean}
  \end{subfigure} \hfill
  \begin{subfigure} {.24\textwidth}
  \includegraphics[width=\linewidth]{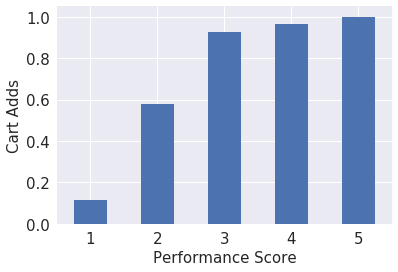}
  \caption{Number of cart adds}
  \label{numCarts_mean}
  \end{subfigure} \hfill
  \begin{subfigure} {.24\textwidth}
  \includegraphics[width=\linewidth]{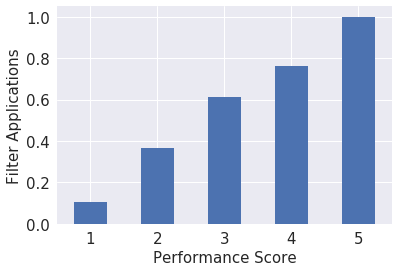}
  \caption{Number of filter applications}
  \label{numFilters_mean}
  \end{subfigure} \\
  \vspace{1pc}
  \begin{subfigure} [t] {.24\textwidth}
  \includegraphics[width=\linewidth]{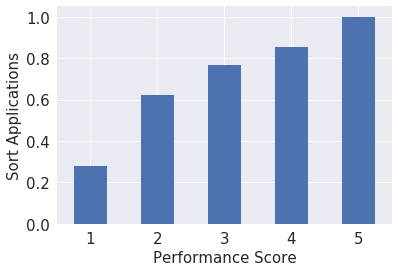}
  \caption{Number of sort applications}
  \label{numSorts_mean}
  \end{subfigure} \hfill
  \begin{subfigure} [t] {.24\textwidth}
  \includegraphics[width=\linewidth]{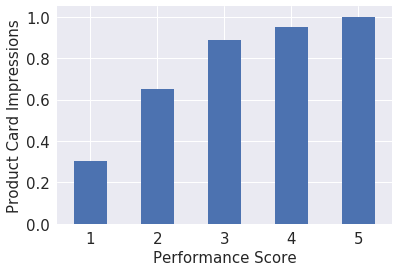}
  \caption{Number of product card impressions}
  \label{numImpressions_mean}
  \end{subfigure} \hfill
  \begin{subfigure} [t] {.24\textwidth}
  \includegraphics[width=\linewidth]{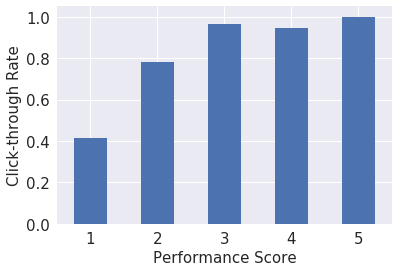}
  \caption{Click-through rate}
  \label{ctr_mean}
  \end{subfigure} \hfill
  \begin{subfigure} [t] {.24\textwidth}
  \includegraphics[width=\linewidth]{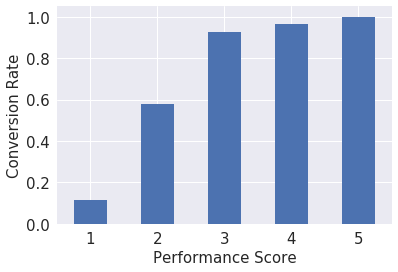}
  \caption{Conversion rate}
  \label{cvr_mean}
  \end{subfigure}
  \vspace{-0.5\baselineskip}
  \caption{Normalized Distribution of Activity Aggregates with respect to Performance Score}
\end{figure*}
\subsection{Experimental setup}
\label{section:experimental_setup}

In this work, we formulate the problem of query performance prediction as a binary classification task, as is done in \cite{productWSDM18}. As described in Section \ref{section:dataset}, we obtained expert judgments for 18,613 queries across a 5-point scale. Similar to \cite{productWSDM18}, we label ``poor,'' ``bad'' and ``average'' queries as DSAT and ``good'' and ``excellent'' as SAT. This results in 6,949 DSAT and 11,664 SAT queries. We treat DSAT as the positive class (classifier target) as the interventions in future based on the model's prediction will be for this class.

We aggregate the metrics, described in previous section, across all the instances of the query in the week to obtain aggregate user behavior corresponding to the query. For metrics which may not have values for all query instances (e.g. timeToFirstClick), we only include instances for which values are present, in the aggregate calculation. These aggregate metrics are used as features for the classification model. We experiment with various descriptive statistics for the features, namely, average, median, standard deviation, inter-quartile range\footnote{In all the figures above, we show qualitative analysis of the features with only the ``averaged'' metric which sufficiently indicates the patterns.}. We bin each numeric feature into 10 percentile buckets and convert them to one-hot encoded features. We also defined certain interaction features such as clickSuccess $\times$ queryCount.

We split the labeled data into 80\% training and 20\% test set. During training, we performed feature selection using recursive feature elimination along with model hyper-parameter tuning. The hyper-parameter tuning is done using five-fold cross validation with class-stratification and optimized for area under the ROC curve (AUC).


\subsection {Results}
\label{section:results}
We analyze the results of our model along the following aspects: performance of learnt classifier, feature importance, performance across e-commerce categories, performance across query types and performance across query volume. We use AUC to evaluate the prediction performance.

\subsubsection{Performance of classifier}
\begin{figure}
  \centering
  \captionsetup{justification=centering}
  \includegraphics[width=\columnwidth]{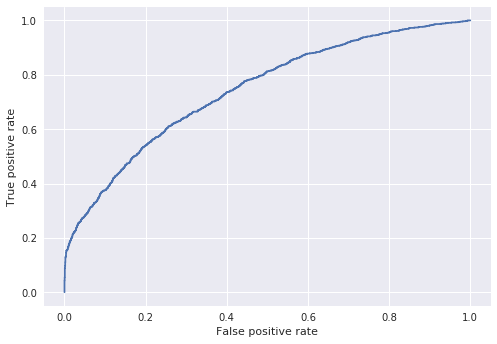}
  \vspace{-1\baselineskip}
  \caption{Receiver Operating Characteristic Curve for Binary Classification of Query Performance}
  \label{fig:auc}
  \end{figure}

\begin{figure}
  \centering
  \captionsetup{justification=centering}
  \includegraphics[width=\columnwidth]{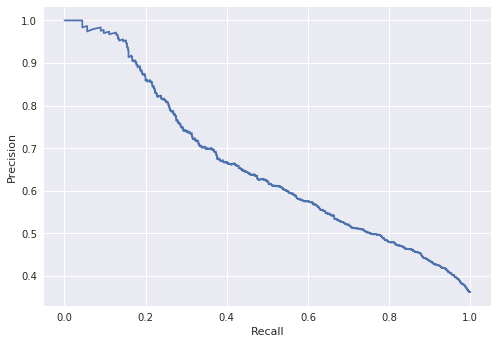}
  \vspace{-1\baselineskip}
  \caption{Precision Recall Curve for Binary Classification of Query Performance}
  \label{fig:pr}
\end{figure}

We train a binary random forest model based on the methodology described earlier in section \ref{section:experimental_setup}. Figure \ref{fig:auc} shows the AUC curve and Figure \ref{fig:pr} shows the PR curve. The overall test AUC obtained is 0.75. We observe that the classifier is able to achieve a reasonably good performance, thus establishing that it is feasible to predict query performance based on user interaction signals.

One application of this predictive model is to enable automated interventions for unsatisfactory queries i.e. when the classifier is confident that the results are poor, we can enable certain interventions like triggering an interactive intent solicitation module. 
Towards that end, we need a reasonably high precision operating point. Based on discussion with business/product team, the operating point that can be used is 85\% precision where we will be able to achieve 20\% recall with the current model.

\subsubsection{Feature importance}
Given below is the list of top-10 most important features based on Gini index: 
\begin{enumerate}
  \item numSwipes
  \item clickSuccess
  \item queryType
  \item wordQueryLen
  \item numProductsFound
  \item cartSuccess
  \item numFilters
  \item numClicks
  \item numSorts
  \item queryCount
\end{enumerate}
We observe a mix of features from various groups in the top features. It is interesting to see the number of page-to-page swipes as a very indicative feature of query performance. We conjecture that the users tend to click and swipe more in exploratory searches when they are satisfied with the initial results and want to continue exploring in the same set without reformulation. As expected, clickSuccess, cartSuccess and numClicks are indicative of query performance. queryType in conjunction with numProductsFound is a good indicator where we expect a small number of products for ``Product'' queries and larger number of products for ``Category'' queries. Interestingly numFilters and numSorts which indicate further refinement of results are also indicative of query performance, where based on Figures \ref{numFilters_mean} and \ref{numSorts_mean} we observe positive correlation with query performance. One surprising observation is that none of the \textit{Activity Time} features are amongst the top 10 features; even though they are indicative, they are less indicative than other structured features like filters and sorts applied.

\subsubsection{Performance across categories}
\label{section:categories}
Table \ref{auroc_cat} shows the performance of the model across the e-commerce categories (described in Section \ref{section:signals}). We observe that the model is able to predict the query performance in ``Mobile'' categories considerably better than all other categories. We conjecture this is due to model's performance across query types (detailed below in section \ref{section:query_types}). The ``Mobile'' category has 7x more ``Product'' queries compared to the ``Lifestyle'' category. Additionally, ``Mobile'' category has 3x less ``Facet Category'' queries. The model is able to perform much better for `Mobile' category due to the underlying query type distribution which is biased towards ``Product'' queries. This is fairly important from a business perspective as the ``Mobile'' category contributes to a significant portion of overall sales.



\subsubsection{Performance across query types}
\label{section:query_types}
Table \ref{table:queryType} shows the results across query types. There are three query types, namely, ``Product,'' ``Facet Category'' and ``Category'' as discussed in Section \ref{section:signals}.

We observe that performance of ``Product'' queries, where the user's intent and language is very specific, is significantly better than other query types. We conjecture that indicators like numProductsFound and numClicks are particularly indicative of the query performance for ``Product'' queries.


\subsubsection{Performance across query volume segments}
Queries are categorized into three segments based on weekly volume: Head, TorsoHigh and TorsoLow. Table \ref{table:queryVol} shows that classifier performance improves as the volume increases. The average queryCount for queries belonging to the Head segment is about 34x that of queries belonging to TorsoBottom segment. Despite the huge difference in amount of data available per query, the classifier is able to predict performance for queries in all three segments reasonably well.






\begin{table}[]
\centering
\caption{Prediction performance for different product categories}
\label{auroc_cat}
\begin{tabular}{|l|l|}
\hline
\textbf{Product Category} & \textbf{AUC} \\ \hline
Books & 0.70 \\ \hline
Electronics & 0.74 \\ \hline
Home And Furniture & 0.72 \\ \hline
Lifestyle & 0.70 \\ \hline
Mobile Phones & 0.90 \\ \hline
\end{tabular}
\end{table}

\begin{table}[]
\centering
\caption{Prediction performance query types}
\label{table:queryType}
\begin{tabular}{|l|l|}
\hline
\textbf{Query Type} & \textbf{AUC} \\ \hline
Category & 0.74 \\ \hline
Facet Category & 0.72 \\ \hline
Product & 0.87 \\ \hline
\end{tabular}
\end{table}

\begin{table}[]
\centering
\caption{Prediction performance for different query volume segments}
\label{table:queryVol}
\begin{tabular}{|l|l|}
\hline
\textbf{Volume Segment} & \textbf{AUC} \\ \hline
Head & 0.76 \\ \hline
TorsoHigh & 0.75 \\ \hline
TorsoBottom & 0.72 \\ \hline
\end{tabular}
\end{table}

%% file: 6-conclusion.tex
Measuring search engine performance is essential to building and improving retrieval algorithms. Query performance evaluation allows for a fine-grained measure of search performance. CTR can be a noisy metric in that high CTR queries may still have poor performance, and vice versa. A more sophisticated analysis of search behavior is needed to distinguish poor and well performing queries. In this work, we successfully demonstrate that query performance can be predicted based on user's interaction with the result set. This is the first study to our knowledge that has collectively defined these signals in the context of query performance prediction for e-commerce search.  Specifically, we propose and use several user interaction signals that help characterize query performance and enabled us to achieve good classification performance using these signals. Notably, our model achieved an overall AUC of 0.75 in the binary SAT/DSAT prediction task.  We have analyzed the results across various factors like category of the query, query type and query volume. Key takeaways from the performance analysis are (a) We achieve significantly higher AUC of 0.90 on certain categories like ``Mobile'' making the result very promising from business impact perspective, (b) Classifier performance varies across query types (``Product'', ``Facet Category'' and ``Category'') and is best for ``Product'' queries, and (c) Classifier performance improves with engagement volume, and is better for Head queries than TorsoBottom queries.

\textit{Future Work} The study can be extended to have a finer prediction target of issue type like spell error, vocabulary gap, selection gap etc. which would make the classifier prediction more easily actionable by giving finer details on the query. Even richer signals of user activities can be used for prediction. For example, the notion of good dwell time (healthy engagement such as reading or voting on reviews) and bad dwell time (unhealthy engagement such as changing seller) might be used. Reducing the number of observations required (currently set to 100) for robustly predicting query performance would be another avenue of future work. This would allow the classifier to scale an even larger number of queries which do not have many instances in a fixed time period.


%% file: 7-acknowledgements.tex
We thank Mr. Priyank Patel and Mr. Subhadeep Maji for their helpful comments.